# #ContextMatters: Advantages and Limitations of Using Machine Learning to Support Women in Politics

**Authors:**
Jacqueline Comer   Sam Work   Kory W. Mathewson   Lana Cuthberson   Kasey Machin

## Abstract

The United Nations identified gender equality as a Sustainable Development Goal in 2015, recognizing the underrepresentation of women in politics as a specific barrier to achieving gender equality. Political systems around the world experience gender inequality across all levels of elected government as fewer women run for office than men. This is due in part to online abuse, particularly on social media platforms like Twitter, where women seeking or in power tend to be targeted with more toxic maltreatment than their male counterparts. In this paper, we present reflections on ParityBOT—the first natural language processing-based intervention designed to affect online discourse for women in politics for the better, at scale. Deployed across elections in Canada, the United States and New Zealand, ParityBOT was used to analyse and classify more than 12 million tweets directed at women candidates and counter toxic tweets with supportive ones. From these elections we present three case studies highlighting the current limitations of, and future research and application opportunities for, using a natural language processing-based system to detect online toxicity, specifically with regards to contextually important microaggressions. We examine the rate of false negatives, where ParityBOT failed to pick up on insults directed at specific high profile women, which would be obvious to human users. We examine the unaddressed harms of microaggressions and the potential of yet unseen damage they cause for women in these communities, and for progress towards gender equality overall, in light of these technological blindspots. This work concludes with a discussion on the benefits of partnerships between nonprofit social groups and technology experts to develop responsible, socially impactful approaches to addressing online hate.

## 1 Introduction

Women continue to be underrepresented at all levels of political leadership around the world (United Nations, 2020a). In fact, only 24 percent of national parliamentarians were women as of November 2018, a small increase from 11.3 percent in 1995 (United Nations,

2020b). This is a real barrier to achieving the United Nations' Sustainable Development Goal of gender equality across all facets of society (United Nations, 2020a) and also a threat to advancing population health (Carles Muntaner and Edwin Ng, 2019). There are fewer women in politics than men, largely because women do not run for office at the same rate as men (Fox, 2001; Lawless and Fox, 2004; Miller, 2016).

This is in part due to harassment on social media (Cuthbertson et al., 2019; Lucina Di Meco and Saskia Brechenmacher, 2020; Rheault et al., 2019; Trimble, 2018). For candidates and elected officials, Twitter is a critical campaign marketing and engagement tool to connect with voters and constituents (Barboni et al., 2018). While women receive the most benefit from social media through visibility and reach, they also experience the worst consequences: a disproportionate amount of harassment because of their gender (Amnesty International, 2018b; Cuthbertson et al., 2019; Tenove and Tworek, 2020).

For example, a recent analysis of the 2020 U.S. congressional races found that female candidates were significantly more likely to receive online abuse than their male counterparts (Guerin and Maharasingam-Shah, 2020). On Facebook, female Democrats running for office received 10 times more abusive comments than male Democratic candidates. Similar trends have been documented in India (Amnesty International UK, 2020a), the UK (Amnesty International UK, 2020b), Ukraine (Otito Greg-Obi et al., 2019), and Zimbabwe (Bardall et al., 2018). This pattern tends to be even more pronounced for female political lead ers from racial, ethnic, religious, or other minority groups (Amnesty International, 2018c); for those who are highly visible in the media (Rheault et al., 2019); and for those who speak out on feminist is sues (Eckert, 2018). The problem is getting worse: toxicity is coalescing into sophisticated and gendered disinformation campaigns targeted at women in public life (Jankowicz et al., 2021).

Contributing to the cycle of gender imbalance in government, women politicians facing online hate feel stress, anxiety, panic attacks, powerlessness, and loss of confidence (Amnesty International, 2018a), making them less effective at their jobs. Further, those women who may be considering advancing their political aspirations, who in turn observe the copious online hate, are deterred from running.

Inspired by the Amnesty International and Element AI called Troll Patrol (Delisle et al., 2018), we built ParityBOT for a grassroots nonprofit organization's efforts to achieve gender equality in politics. In 2019, this organization's leaders anecdotally shared in media coverage that nearly every conversation with women considering running for office included hesitation due to online abuse (Mertz, 2019).

ParityBOT is a Twitter bot that uses machine learning, specifically natural language processing (NLP), to detect toxic tweets directed at women candidates during an election period. For each tweet that passes a predetermined toxicity threshold, the bot posts a positive tweet. We deployed ParityBOT before, during, and over several days following: 1) the 2019 Canadian federal election (September 23 - October 26), the 2020 New Zealand general election (August 12 - November 4), and the 2020 U.S. presidential election (September 28 - November 17).  In both the Canadian and US elections, the bot posted a positivitweet on its own Twitter account. Based on initial results, a design decision was made in the New Zealand election to send a randomly selected positive tweet to a candidate directly if the bot was triggered by any toxic tweet.  The idea behind this tool is to go beyond measurement and into direct intervention in a digital community to promote a shift in our cultural vocabulary ([Marantz, 2019](#)). In this case, the cultural intervention aims to improve the environment for current and future women political candidates, and ultimately, provide equal opportunities for women to enter politics by removing the barrier to gender equality that a toxic Twitter environment creates.

We collected and analyzed more than 12 million tweets directed at women candidates across the three elections in three different countries. In each of the elections, there was a notable pattern in terms of which women received the most harassment on Twitter: the higher the candidate's profile, the more abusive tweets they received. The most targeted women in each election accounted for an overwhelming percentage of the total abuse. However, while our data shows that toxicity towards women candidates is prolific, due to model constraints (i.e., Twitter limiting how often our bot could tweet) our results only show a fraction of the real problem. Further, our ParityBOT model also failed to detect as "toxic" some well-documented insults used to target those specific candidates. Below are three case studies about Catherine McKenna (Canada), Jacinda Ardern (New Zealand) and Kamala Harris (United States). The results highlight the limitations of NLP-based models. We discuss future research directions in NLP towards models that can better detect both overt and covert toxicity. These results serve to raise awareness and to continue to build interventions to help tackle online abuse towards women election candidates, and help reach gender parity in politics worldwide.

# 2 Methods

In this section, we outline the technical details of the original ParityBOT system ([Cuthbertson et al., 2019](#)).

## 2.1 ParityBOT System

At a high level, ParityBOT "listens" to Twitter for mentions of a list of queries (i.e., Twitter handles of women candidates running for office in a given election) using the open-source Python library Tweepy. When a query of interest is mentioned, ParityBOT cleans and analyses the tweet, and stores the tweet and features in a datastore. Cleaning is done following a predefined list of text processing rules, and analysis is done using a text analysis model described below. If a tweet is predicted by the system to be toxic, ParityBOT posts a positive tweet sampled from a predefined list with tweets written by subject matter experts, and crowdsourced from engaged audience members. The system was used only for English language tweets, but it is possible to expand beyond English in the future. Data is collected and used according to acceptable terms of use as outlined in Twitter's Developer Agreement and Policy ([Twitter Developer, 2020](#)).

ParityBOT scores the tweet using a text analysis model called Perspective API[1], which includes 11 classification categories and uses a machine learning system to provide scores for each classifier. ParityBOT pays particular attention to one of these classifiers and its score, TOXICITY, to establish a decision threshold for whether the bot posts a positive tweet. No user features are included in the analysis of the tweets in order to avoid any potential bias.

## 2.2 Twitter Handles, Candidate Gender and Writing Positive tweets

In order to create a list of each woman candidate's Twitter handle, a group of volunteers created a candidate list or used candidate lists provided online by political parties, community groups, post secondaries, or nonprofit organizations ([Cuthbertson et al., 2019](#)). They used online searches and Twitter's search interface to find the most likely handle for each candidate. Volunteers used their best judgement based on research to confirm the gender identity of candidates. In the Canadian election, to predict the gender of each candidate based on their first names, a Python library gender-guesser was used ([Cuthbertson et al., 2019](#)). To attempt to confirm gender predictions, we used manual validation. In the New Zealand election, all the women candidates were collected directly from publicly available sources on the internet. In the U.S. election, we used a publicly available list from the Center for American Women in Politics from Rutgers University[2] to compile the candidate list, and volunteers filled in each Twitter handle by searching candidate names on Twitter's built-in search engine.

Volunteers wrote [positive tweets](#) (called "positivitweets") for ParityBOT to post, and as ParityBOT was used for more elections, the bank of positive tweets grew and was edited to reflect local context and languages (for example, French and English in Canada, and Te Reo Maori and English in New Zealand). Positivitweets were also crowdsourced through a Google form posted to the ParityBOT Twitter page, asking engaged audience members to submit a positive tweet for the bot to tweet. We vetted and edited these submissions to ensure a tonal match with the uplifting, informative, and factual pre-existing bank of positive tweets.

---

[1] Perspective API can be found at [http://www.perspectiveapi.com](http://www.perspectiveapi.com). Due to privacy and sensitivity concerns, ParityBOT data is only available upon request.

[2] [https://cawp.rutgers.edu/](https://cawp.rutgers.edu/)

| Bot | Positivitweets in library |
|---|---|
| ParityBOT_CA (Canada 2019) | 227 |
| ParityBOT_NZ (New Zealand 2020) | 141 |
| ParityBOT_US (United States 2020) | 539 |

Table 1: Number of positivitiweets in each bot library. This data does not include the crowdsourced tweets collected throughout each election cycle.

## 3 Results and Outcomes

ParityBOT's NLP model analyzed a total of 12,688,346 tweets sent to women candidates from the 2019 Canadian federal election, 2020 New Zealand general election and the 2020 U.S. presidential election. For each tweet collected, the ParityBOT model calculated the probability that the tweet was abusive or hateful, as defined by its TOXICITY score. A positive tweet was posted by ParityBOT if the probability that a tweet was toxic was greater than the set response decision threshold. This decision threshold differed slightly across each election due to the number of positive tweets in our library, the volume of tweets to process, and the Twitter API daily limit ([Cuthbertson et al., 2019;](#) [Twitter Developer, 2020](#)).

For the Canadian federal election, we set the initial decision threshold to 0.8, but then increased it first to 0.9 and again to 0.95 to reduce the number of positive tweets we posted per day to manage the experience of someone following ParityBOT. Using this threshold pattern, our model classified 3,381 tweets as hateful, and posted 2,428 positive tweets. We used the 0.9 threshold for ParityBOT_NZ, and the model classified 4,575 tweets as hateful, and sent out 3,906 tweets. For ParityBOT_US, the decision threshold was set to 0.99 due to the increased number and rate of tweets processed, considering the number of U.S. adults who use Twitter ([Perrin and Anderson, 2019](#)). Following ParityBOT on Twitter has to be a manageable experience, and Twitter restricts the frequency at which a bot can tweet. Of the more than 12 million tweets analyzed across the three elections, 20,411 were scored "toxic" based on these thresholds, and the bot sent out 19,537 positive tweets. A positive tweet was not sent out for every classified hateful tweet, which reflects the decision rate-limit of ParityBOT ([Cuthbertson et al., 2019](#))

| Election | Days bot in operation | Toxicity decision threshold | Candidates tracked | Tweets analysed | Tweets classified as hateful | Positive tweets sent |
| --- | --- | --- | --- | --- | --- | --- |
| Canada 2019 | 33 | > 0.9 | 90 | 228, 255 | 3, 381 | 2, 428 |
| New Zealand 2020 | 84 | > 0.9 | 86 | 199, 578 | 4, 575 | 3, 906 |
| United States 2020 | 50 | > 0.99 | 376 | 12,247,817 | 12, 455 | 12,230 |
| Total | 167 | N/A | 1, 251 | 12,688,346 | 20, 411 | 19, 537 |

Table 2: ParityBOT Tweet Data Across Three Elections. Note: Due to a COVID-19 lockdown in New Zealand in August 2020, on August 17 the New Zealand election was moved from September 17 to October 17, extending the election period by one calendar month. We chose to keep ParityBOT_NZ running over the extended campaign period.

Across the three elections,1,251 Twitter users followed the bot accounts, including 38 candidates, some of whom engaged with and thanked the bot for supporting them. The bots also garnered mainstream media attention, which helped raise awareness of the issue of online abuse targeting women candidates, and brought the idea of using machine learning for social good to a wider, non-scientific audience. Further, analysis of ParityBOT_NZ data shows that over the New Zealand election period, the bot had 247,684 impressions and an engagement rate of 2.3%, significantly higher than the median Twitter engagement rate across every industry of 0.045% ([Feehan, 2021](#)).

## 4 Model Limitations and Discussion

There are several important limitations of the ParityBOT model as outlined in [Cuthbertson et al., 2019](#). While assigning a candidate's gender identity was automated and then manually verified, there still exists inherent uncertainty in the absence of an official source to confirm with. Additionally, our system does not currently detect non-binary gender, however to the best of our knowledge, based on cross referencing with self-identification in online content, we did not identify any non-binary candidates. Furthermore, without a human in the loop at the final decision stage, using an NLP model to identify harmful content cannot guarantee accuracy. The tradeoff between the benefits of automating such decisions to track and filter at scale and the current limitations of reliability are well known ([Natasha Duarte et al., 2017](#)). To investigate the performance of ParityBOT during its election deployments, we analyzed the results for false positives and contextualized false negatives.

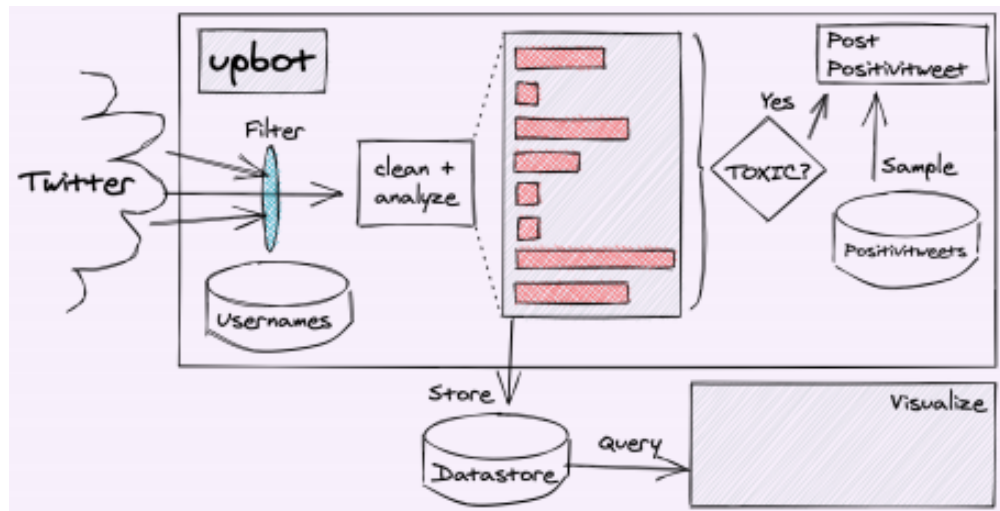

Figure 1: ParityBOT Overview. This diagram shows the components of the ParityBOT system specifically illustrating data ingest from Twitter, to cleaning and analysis, and potentially the posting of a positive tweet. There are three databases for usernames, positive tweets, and temporary storage for the stored featurized data for visualization.

## 4.1 False Positives

One way to assess a classifier's quality is by looking at its false positive and false negative rates. The consequences of ParityBOT misclassifying tweets as false positives may result in the bot tweeting too frequently and unnecessarily, while a failure to catch true toxic tweets under-reports the magnitude of the problem of toxicity, undermining the usability of the bot. One concern was that the high confidence thresholds (>= 0.9) chosen for the three elections to determine tweet toxicity resulted in downplaying the true scope of the problem faced by women candidates. This problem was compounded, as tweets have additional content, such as pictures, links and other media, that would not have been included in the tweet text. This was not analyzed by the model, potentially hampering its classifying ability. In short, analyzing false positives is important to better understand deployment issues for future ParityBOT iterations, especially when considering other interventions and extending the usage of the system.

Previous literature has investigated the false positive rate of toxicity detection models as a potential source for bias against minority groups ([Dixon et al., 2018](#), [Zhou et al., 2021](#)). In particular, curse words often triggered high toxicity scores despite often neutral or even positive intentions in usage on social media.

To determine the rate of false positives, from the total of 4,575 tweets rated by the ParityBOT_NZ as hateful, we randomly selected 355 tweets to be manually classified. This sample of tweets was divided among three human reviewers to classify as "toxic" or "not toxic". Instructions for what constituted toxicity in a tweet were minimal, with the reviewers simply being told they should feel quite certain a tweet was toxic when labelling it as such. This was an intentional approach meant to approximate the real-world settings, where toxicity is, to a certain extent, subjective in interpretation, and to mimic the high probability threshold at which ParityBOT had been set to define toxicity. Furthermore, we assume the average user on Twitter is unlikely to have any sort of expertise or training in toxicity recognition, but simply encounters these tweets in the wild, so to speak. For our human annotators, only text tweet content was provided, which included hashtags, URLs and other users mentioned, but not emojis, pictures, or other media (ParityBOT itself only parses text).

Each reviewer was given 100 unique tweets from the sample to label independently. To assess inter-labeller agreement, all three reviewers also independently classified the same set of 55 tweets as toxic or not, and Fleiss' kappa was calculated ([Fleiss, 1971](#)). Based on this, inter-labeller agreement could be considered only "fair", with a Fleiss' kappa value of 0.22. Where there was disagreement amongst reviewers, the ground truth was determined by majority (two out of three).

Of the 55 sample toxic tweets, 60% were determined to be "true" toxics and 40% were false positives. From the entire sample of 355, 184 were labelled "toxic" and 171 "not toxic".

| @ParityBOT_NZ Toxic > 0.9 | Human Labelled Toxic | Human Labelled Not Toxic |
|---|---|---|
| 335 | 184 | 171 |
| Percentage of toxic tweets | 52% | 48% |

Table 3: Sampled New Zealand False and True Positives

These results were somewhat surprising. The lack of strong inter-labeller agreement along with the rate of "true" and "false" positives suggest inherent issues with the toxicity classifier. For comparison, Breitfeller et al. ([2019](#)) also used three human annotators on a set of 200 posts and achieved a Fleiss' kappa of 0.424, considered a 'moderate' level of agreement, but acknowledged the difficulty of the task and lack of objectivity involved. The tweets for which two out of three annotators agreed were false positives (i.e. "not toxic") from the inter-labeller agreement sample is given in Appendix 7.2.

Given that perceptions of 'toxicity' are highly subjective and may be influenced by a myriad number of factors, including race, ethnicity, gender and class, it is important to recognize that the characteristics of the three human annotators here were relatively homogenous ([Sap et al., 2019](#)). All three were caucasian, two women, one male, and of relatively similar class and educational background. Assuming this relative homogeneity amongst the reviewers, the disparate agreement results are even more interesting. Further investigation is warranted to investigate the discrepancy between NLP classifiers and what may constitute a ground truth of "toxicity" by human labellers on a corpus of tweets in this domain.

## 4.2 False Negatives and Microaggressions

One significant limitation of our model is that Perspective API (and other similar sentiment analysis tools) was not designed to surface subtly hateful, contextual tweets or microaggressions, as defined by [Derald Wing Sue (2010)](#) as "everyday verbal, nonverbal, and environmental slights, snubs, or insults, whether intentional or unintentional, which communicate hostile, derogatory, or negative messages to target persons based solely upon their marginalized group membership".

Perspective API uses crowd-sourced labels in their training data of whether a comment is toxic, and defines toxic as "a rude, disrespectful or unreasonable comment that is likely to make people leave a discussion" ([iislucas, 2018](#)). However, toxicity is not always expressed with toxic language. In fact, many microaggressions use positive language, and yet to the groups on the receiving end of the comment it is no less offensive than a more obviously negative slight [(Jurgens et al., 2019;](#) [Field and Tsvetkov, 2020](#)). Context matters, and because microaggressions are subtle and often unconscious, toxicity is in the eye of the beholden.

Furthermore, these microaggressions can be crouched in subtle or seemingly innocuous terms. [Marguerite Ward and Rachel Premack (2018)](#) note that if "the words look and sound complimentary, on the surface (they're most often positive), [the receiver] can't rightly feel insulted and doesn't know how to respond". The insidiousness of the microaggression is found in its duplicitous nature—on one side, you have a snub or insult that is meant to diminish the receiver based on belonging to a marginalized group, and on the other side you have self-doubt and questioning of the receiver's perception of reality, and whether they have the right to be insulted and speak up.

To contextualize how microaggressions allow users with harmful intent to bypass an automated system looking for more obvious signs of directed abuse, we examine three case studies taken from ParityBOT (Canada), ParityBOT_NZ and ParityBOT_US data. These look at both overt and covert microaggressions targeted at Minister Catherine McKenna, Prime Minister Jacinda Ardern and Vice President Kamala Harris, respectively. We chose these case studies because they represent the three women candidates who received the most negative tweets and a very high proportion of total negative tweets across the three elections based on our system. These case studies represent specific examples of the general limitations of our model to surface any or all microaggressions based on our model application.

**"Climate Barbie" - Catherine McKenna Case Study**
Catherine McKenna was elected to Canadian parliament in 2015, and served as Minister of Environment and Climate from 2015–2019. In 2017, another Canadian member of parliament referred to McKenna as "climate Barbie'' on Twitter, to which McKenna replied, "…We need more women in politics. Your sexist comments won't stop us" ([CBC News, 2017](#)). While the offending MP apologized the following day in parliament, the name stuck, and in 2021 the moniker is still used on Twitter to deride McKenna, despite McKenna not being the environment minister since November 2019. While "Climate Barbie'' is overtly sexist, it does not meet the "toxic" threshold of the Perspective API NLP model used in ParityBOT, and therefore is classified as a microaggression. The phrase "Climate Barbie '' scores a 0.0597 for TOXICITY using Perspective API's model, compared with the thresholds above 0.8 used to determine TOXICITY in ParityBOT.

Catherine McKenna received the highest percentage of toxic tweets (33%) sent to women candidates during the Canadian election. In fact, it was due to the high number of tweets specifically targeting McKenna that the decision threshold was increased from > 0.8 to > 0.9. There were 471 tweets containing "Climate Barbie" and 24 of those were labelled as "toxic". The model failed to classify 447 instances of "Climate Barbie" as toxic, because of the nuanced nature of the "Climate Barbie" microaggression and Perspective API's inability to surface contextualized hateful speech.

| Total analysed tweets | 228,225 |
|---|---|
| Total tweets >0.9 likely to be toxic | 3,381 |
| Total tweets to @cathmckenna | 50,063 |
| @cathmckenna tweets >0.9 likely to be toxic | 1,113 |
| Percentage of total toxic tweets sent to @cathmckenna | 33 |
| Tweets containing "Climate Barbie" | 471 |
| "Climate Barbie" tweets >0.9 likely to be toxic | 24 |
| Climate Barbie"-related false negatives | 447 |

Table 4: "Climate Barbie" Tweet Data - ParityBOT (Canada)

### "Cindy" - Jacinda Ardern Case Study

Jacinda Ardern became Prime Minister of New Zealand in 2017, and was up for reelection in 2020. Her critics often call her "Cindy", a paternalistic name that Ardern has said she hates ([Dowd, 2018](#)) and is used to belittle her, to point out that "…no matter what she achieves, handles, or leads, they want to…remind her that she is still just a girl" ([Vance, 2020](#)). In short, "Cindy" is a microaggression.

Throughout the election period, Ardern received 54% of the total number of toxic tweets sent to women candidates in the 2020 New Zealand election. ParityBOT_NZ collected 685 tweets referencing "Cindy", of which only 24 were scored as "toxic". This means our model "missed" at least 661 potentially hateful tweets.

| | |
|---|---|
| Total analysed tweets | 199,578 |
| Total tweets >0.9 likely to be toxic | 4,575 |
| Total tweets to @jacindaardern | 95,915 |
| @jacindaardern tweets >0.9 likely to be toxic | 2,475 |
| Percentage of total toxic tweets sent to @jacindaardern | 54% |
| Tweets containing "Cindy" | 685 |
| "Cindy" tweets >0.9 likely to be toxic | 24 |
| "Cindy"-related false negatives | 661 |

Table 5: "Cindy" Tweet Data - ParityBOT_NZ

### "Commala" - Kamala Harris Case Study

Kamala Harris was elected U.S. Vice President in October of 2020, becoming the first woman and first person of colour to hold the office of the vice presidency. During her campaign, her opponents and some members of the media continually mispronounced her name, and persistently used the term "Commala", a form of casual yet malicious racism; as such, "Commala" is a microaggression. This type of seemingly small aggression is known to accumulate over time, and cause feelings of exclusion and unworthiness of respect ([Jankowicz et al., 2021](#)). This sparked a backlash with supporters (mainly BIPOC people) responding with the trending hashtag #MyNameIs ([Flynn, 2020](#)).

Harris received 28% of the total number of "toxic" tweets sent to women candidates in the

2020 US election. Between September 1 and November 1 2020, ParityBOT_US collected 589 tweets referencing "Commala", of which only 14 were scored as "toxic". This means that our model missed classifying at least 575 potentially hateful tweets.

| Total analyzed tweets | 12,247,817 |
| --- | --- |
| Total tweets >0.9 likely to be toxic | 348,315 |
| Total tweets to @KamalaHarris | 2,430,184 |
| @KamalaHarris tweets >0.9 likely to be toxic | 97,330 |
| Percentage of total toxic tweets sent to @KamalaHarris | 28% |
| Tweets containing "Commala" | 589 |
| "Commala" tweets >0.9 likely to be toxic | 14 |
| "Commala"-related false negative | 575 |

Table 6: "Commala" Tweet Data - ParityBOT_US. Note: We set our data query decision threshold here to > 0.9 for consistent comparison across the Canadian, N.Z., and U.S. elections.

It is probable that our model missed other microaggressions directed at women candidates across all three elections, due to the aforementioned limitations of Perspective API. However, this issue is not only a problem with Perspective API but also with other widely used NLP models—they are not designed to detect veiled, contextual toxicity ([Breitfeller et al., 2019](#); [Field and Tsvetkov, 2020](#); [Jurgens et al., 2019](#)). This is because: 1) current technology focuses on overt abusive language and hate speech, without considering the different forms that abuse can take ([Jurgens et al., 2019](#)), 2) labelled microaggressions are not represented in existing NLP toxicity datasets ([Han and Tsvetkov, 2020](#)), and 3) there are inherent challenges with identifying coded abuse, with the risk that crowd-sourced labels may obscure important minority perspectives ([Natasha Duarte et al., 2017](#)).

In the UNDP's Human Development Report ([2019](#)), the fight for gender equality is framed as part of a larger struggle against systemic power imbalances. This power gap is also evident in our NLP data models. If our NLP models come from power structures of the past and present that disadvantage marginalized communities like women seeking elected office, and we use training data with inherent and unconscious biases that maintain those power structures, then we cannot rely on current machine learning tools alone to change the future. However, by using tools like ParityBOT to publicize the issue of online harassment and its effects on the power imbalance in our political systems between men and women, we can begin to support and empower the unempowered to have equal seats—and voices—at the table. It cannot be understated that being able to focus on these issues and work towards solutions that identify and solve technological blindspots for underrepresented communities, and develop responsible, socially-impactful approaches to addressing online hate, is a significant benefit of partnerships between nonprofit social groups and technology experts.

## 5 Future Research and Conclusions

We would like to explore building NLP models that can surface both macroaggressions and microaggressions, recognizing that not only is deep semantic analysis difficult, but it is also extremely challenging given the very nature of microaggressions: they are subjective and often unconscious, making data mislabeling a real issue, thanks to the annotators' own biases ([Han and Tsvetkov, 2020](#)). Even when assessing the reliability of ParityBOT_NZ, our human labellers had only fair agreement as to what constituted a truly toxic tweet, and consequently their perceptions of toxic tweets were far from consistent, highlighting the subjective nature of interpretation on an individual level. Do systems such as ParityBOT, which aggregate individual human experience at a massive level, achieve some feasible universality of experience? Or do they conceal minority opinions, further marginalizing legitimate experiences? Further work could assess the role context plays in the interpretation of tweets, such as whether being able to see the tweet in situ as a thread of replies and with the additional con tent of pictures, links, emojis, and even Twitter user bios that are missing from the ParityBOT anal ysis. Technical methods of mitigating NLP system labelling biases will be another investigation of future work, such as using unsupervised learning to surface implicit gender bias in text ([Field and Tsvetkov, 2020](#)) and employing fairness constraints to avoid unintended bias ([Gencoglu, 2020](#)).

ParityBOT highlights that hate speech towards women election candidates, especially those in or seeking high-profile positions of power, is prevalent on social media platforms, which affects not only gender equality in politics, but also the demo cratic and overall health of our communities ([Carles Muntaner and Edwin Ng, 2019](#)).

We know that ParityBOT makes a positive difference for women seeking office, and while partner ing nonprofit groups and technology experts makes positive impacts for addressing online hate, there is still work to do to adapt our NLP models to effectively identify both overt and covert toxicity to evolve the discourse and affect real change to reach gender parity in politics.

## 6 Impact Statement

The global gender and power imbalance in political office is also reflected in the online abuse that went both detected and undetected by our NLP model. NLP models like the one used with ParityBOT are trained on corpora which reflect the language of the empowered majority—patriarchal, predominantly hetero-normative, and Caucasian in Western societies—enshrining and promulgating their perceptions and value judgments to the disadvantage of marginalized communities. To effectively use NLP to address online hate towards women seeking power, and by extension change society, our models need to reflect a future we want rather than a past we reject, and while partnering technology experts with nonprofit organizations is a great step forward to reaching gender equality, the onus of change must be on those who hold the power—i.e., social media platforms and technology experts—and not those seeking it.

# 7 Appendix

## 7.1 Sample Positivitweets in Bot Libraries

**ParityBOT_CA**

- « Agissez toujours comme s'il était impossible d'échouer » Winston Churchill
- 2018: Québec is the first jurisdiction to elect more than 40% women (41.5%)
- Keep using your voice! Our democracy relies on diverse voices being heard
- 1993: Kim Campbell is Canada's first (and still only) woman Prime Minister
- Thank you for putting yourself out there to serve. Credit: @rachelmcgraw11
- We need you. Your voice and your experiences are valuable and important.
- Shattering ceilings and fragile egos in one fell swoop. #womeninpolitics
- Keep up the great work! We need more women like you in political office!
- 1917: Louise McKinney is the first woman elected in Canada (Alberta MLA)
- Women should be at all decision making tables. #addwomenchangepolitics
- 1991: Rita Johnson is first woman Premier in Canada (British Columbia)
- Don't be afraid. Be focused. Be determined. Be hopeful. Be empowered.
- « Je ne perds jamais. Soit je gagne, soit j'apprends » Nelson Mandela
- You are inspiring young women to have their voices heard! Keep going!
- Square your shoulders. Face the sun. Be brave. The world needs you.
- You're stronger than you know! Keep fighting for what you believe

**ParityBOT_NZ**

- "I never, ever grew up as a young woman believing that my gender would stand in the way of anything I want." - Jacinda Ardern.
- "I respect myself and insist upon it from everybody. And because I do it, I then respect everybody, too." - Maya Angelou
- "I refuse to believe that you cannot be both compassionate and strong." - Jacinda Ardern #nzpol #NZPolitics
- "Listen, you can think anything of me but what I'm going to show you is what I can do and if you're not happy, then I don't really care." - Dominique Crenn

- Mauria te pono
- When you put in the hard mahi, good things happen. #nzpol #NewZealand #NZPolitics #womeninpolitics
- He ora te whakapiri; He mate te whakatiriri
- He whai nui kei runga noa ake
- Aroha ki te tangata; Ahakoa ko wai te tangata
- You've done the hard yakka. Keep going!
- He aha te kai ō te rangatira? He Kōrero, he kōrero, he kōrero.
- "People do not decide to become extraordinary. They decide to accomplish extraordinary things." - Sir Edmund Hillary
- 101 years ago, Kiwi women gained the right to stand for parliment. That's 101 years of great women like you running to make NZ a better place. #nzpol #NewZealand #NZPolitics #womeninpolitics
- She'll be right. She'll always be right.
- It's not the mountains we conquer but misogyny. #nzpol #NZPolitics
- #grit&grace
- From dissension, envy, hate...you make our country good and great
- Peace not war shall be your boast; But should foes assail your coast, Show them you're a mighty host. #nzpol #NewZealand #NZPolitics #womeninpolitics

**ParityBOT_US**

- If you want to lift yourself up, lift up someone else. - Booker T. Washington
- Roses are red, violets are blue. You've joined a tough race. We're so proud of you. #womeninpolitics
- Why do we need more women in politics? To improve our democracy and encourage equal representation.
- If you wouldn't go to someone for advice, don't pay heed to their opinions, either. You've got this.
- Women from many backgrounds, ethnicities and political leanings should step up and run for office.
- The more women who run for office, the more doors get opened for even more women to run for office.

- You are making a difference, and are a role model to young women everywhere. Keep up the good work!
- Anyone can ask a woman to run for office. Women need to be asked several times. #askher to run!
- Thank you for being a candidate! We want more female role models for girls. Good luck!
- Her gender is irrelevant. What's relevant is her courage to stand up and run.
- You are an inspiration to young women and girls everywhere! Keep working hard! We support you!
- We need to not only have women run, but win. Others need to go out and show support by voting!
- Women, you have power and grit and can get it done--run for office. #addwomenchangepolitics
- Every vote makes a difference. Your voice makes a huge impact!
- "My persuasion can build a nation. Endless power. Who run the world? Girls!" - Beyonce
- Continue to be strong, believe in yourself and you will succeed. Ignore the naysayers!
- To #womeninpolitics: you're stronger than anyone who criticizes you.

## 7.2 Tweets Labelled "Not Toxic" (False Positives) By Two Out of Three Human Annotators

- @Rondaful78 @NathanF_NZ @PouTepou @paulabennettmp Well you are talking shit then.
- @CattFalstaff @WhiteHouse @realDonaldTrump @SecAzar @soychicka üî•‚ÄºÔ∏èWe know damn well that so-called-Putin-vaccine for COVID will kill pple, which is like taking polonium @ICCT_TheHague @NATO @WHO @UNHumanRights @Refugees @JustinTrudeau @jacindaardern @ScottMorrisonMP @NewYorkStateAG @SpeakerPelosi @RepAdamSchiff @RepKatiePorter @SenWarren
- @bleachley @Megan_Woods Are you seriously that stupid??????
- @Arthur89367695 @nzherald Finally these idiots have a plan ... After how many months? @jacindaardern @AshBloomfield #nzpol

- @JudithCollinsMP you're embarrassing yourself. The more you critise the PM, the more you look like an idiot. Bridges did the same thing and look what happened. Why don't you talk about what national can actually do for the country in the future. Also there's no government cover-up. Wake up.
- @paulabennettmp National supporters aye. Such dumb arses
- Fuck off @jacindaardern @GrantRobinsonMP you will not destroy us.. WE WILL REVOLT AND SAY #NoLockdown #NoMasks
- @JGFireHorse1 @SevenSharp @jacindaardern #BilaryHilary is stupid. #nzpol
- @emmamellow Why are New Zealanders so dumb ? ü§îü§îü§î
- @jacindaardern time to kick Winston's boney ass!!!
- @Iminwellington @jacindaardern Leave poor Winnie's arse alone. He had to spend all last lockdown horse riding.
- @carol_stirling @KrittiAKL @SiouxsieW @jacindaardern I know many people just like that. In NZ and overseas. I find it mind blowing but ultimately I think this is driven by fear and exceptionalism. Or maybe these people are just fucking stupid despite their university degrees ü§®ü§Ø

## 7.3 Sample Tweets Sent To Candidates Using Targeted Microaggressions

### 7.3.1 Tweets (toxicity score) directed at Catherine McKenna

- Well Climate Barbie that is another big promise you have zero chance of seeing through as you are done October 21. Don't worry many of us will give a fine send off! (0.192)
- Hey. Climate Barbie. I thought we were fucked in 12 years? Is this something you're going to say loud. Say often. And say in parliament, hoping that everyone will believe you? Is it Climate Barbie? (0.714)

  To think you've been in your position for 4 years and have accomplished nothing says alot Climate Barbie. But than again you're just dumb plain and simple. . . (0.782)

- Where's @cathmckenna aka Climate Barbie aka Poo Barbie aka Sewage Hag aka Sewage Barbie been hiding? Would have thought she'd been with Greta and Trudeau for the Photo Op. (0.322)
- Let's hope we can do this let climate barbie get into her barbie car and drive off into her barbie world never to be see again. (0.381)
- To think that in 15 months, a mentally challenged girl could lie to everyone around the world with that scowl demanding everyone take action except herself. Never

underestimate the power of one girl to lie.  Actually 2.  Climate Barbie Fixed it. Thank me later (0.246)

### 7.3.2  Tweet (toxicity score) directed at Jacinda Ardern
- cindy is in bed with the ccp no doubt about it (0.274)
- coz auntie cindy paid them off too they know where their breads buttered fakenews (0.175)
- years ago i went to madam tussauds now that is freaky to have a valid dinner convo on the cindy hot or not topic the best way would be a life sized cindyquin at the table may put most off their meal though except commander dc (0.185)
- hey super star cindy is the communist sympathetic socialist (0.203)
- cindy stfu please (0.568)
- data doesnt lie but cindy does (0.281)

### 7.3.3  Tweet (toxicity score) directed at Kamala Harris
- naw heifer i wont send you any money a good democrat would steal it commala harris extreme  liberal  imagine  if  biden  is  unable  to serve this spiteful woman could use our constitution to employ the 25th amendment to elevate herself to become president vicious (0.515)
- Comma-la is a joke.  Even more unlikeable than Shrillary. There's a reason they kept her off the campaign trail!" (0.690)
- Your an absolute joke comma-la heels up Harris! (0.650)
- Say, that reminds me, did you-all meet Kamala? Huh? What? Camel-uh or Comma-la? Huh? That reminds me, what's that Tony Soprano's wife's name?  Yeah?  I like caramel corn. Say, that reminds me, I like kids jumping up on my lap. (0.325)
- The perky donkey is a problem for everyone, her party is petrified of her.  Biden will fall on his arthritic knees to every radical .AOC demand. Joe's too weak to resist. Comma-la will jump to enact any ideological policy the squad wants. https://t.co/rcDbIPAk9s (0.473)
- Would you prefer Commala or Commyla? Asking for a friend. (0.057)
- So  it's  [camel]-A?  Clamata is like the olive. Comma-la? Maybe just #heelsupharris (0.068)